\documentclass[fleqn,useAMS,usenatbib]{mnras}
\usepackage{graphicx}	
\usepackage{lineno}
\usepackage{color,soul}		
\usepackage[T1]{fontenc}
\usepackage{graphicx}	
\usepackage{amsmath}	
\usepackage{amssymb}	

\title[A long term multi-frequency study of solar rotation]{A long term multi-frequency study of solar rotation using solar radio flux and its relationship with solar cycles}

\author[Vivek Kumar Singh, Satish Chandra, Sanish Thomas, Som Kumar Sharma \& Hari Om Vats]{Vivek Kumar Singh,$^{1}$\thanks{vivek.singh@shiats.edu.in (VKS),} Satish Chandra,$^{2}$\thanks{satish0402@gmail.com (SC),} Sanish Thomas$^{1}$\thanks{sanish.thomas@shiats.edu.in (ST),} Som Kumar Sharma,$^{3}$\thanks{somkumar@prl.res.in (SKS),} and Hari Om Vats$^{4}$\thanks{serfahmedabad@gmail.com (HOV)}\\
$^{1}$Department of Physics, Sam Higginbottom University of Agriculture Technology and  Sciences, Praygraj, India\\
$^{2}$Department of Physics, Pt. Prithi Nath PG College, Kanpur, India\\
$^{3}$Physical Research Laboratory, Ahmedabad, India\\
$^{4}$Space Education and Research Foundation, Ahmedabad, India}

\begin{document}

\date{Accepted XXX. Received YYY; in original form ZZZ}

\pagerange{\pageref{firstpage}--\pageref{lastpage}} 

\pubyear{2020}

\maketitle


\label{firstpage}

\begin{abstract}

This paper examines long-term temporal and spatial fluctuations in the solar rotation (more than four solar cycles) by investigating radio emission escapes from various layers of the solar atmosphere during the years 1967-2010. The flux modulation approach can also be used to investigate variations in solar rotation, which is a contentious topic in solar physics. The current study makes use of a time series of radio flux data at different frequencies (245-15400 MHz) obtained at Sagamore Hill Solar Radio Observatory in Massachusetts, USA, and other observatories from 1967 to 2010. The periodicity present in the temporal variation of time series is estimated through Lomb Scargle Periodogram (LSP). The rotation period estimated for five radio emissions (606, 1415, \& 2695 MHz; from corona, and 4995 \& 8800 MHz; from transition region) through statistical approach shows continuous temporal and spatial variation throughout the years. The smoothed rotation period shows the presence of $\sim$ 22-yrs periodic and $\sim$ 11-yrs components in it. The 22-year component could be linked to the reversal of the solar magnetic field (Hale's) cycle, while the 11-yrs component is most likely related to the sunspot (Schwabe's) cycle. Besides these two components, random components are also prominently present in the analyzed data. The cross-correlation between the sunspot number and the rotation period obtained shows a strong correlation with 11-yrs Schwabe's and 22-yr Hale cycle. The corona rotates faster or slower than transition region in different epoch. The swap of faster rotation speed between corona and transition region also follows the 22-yrs cycle.

\end{abstract}


\section{Introduction}

From long radio waves to short X-rays, the Sun releases radiation across a wide variety of energy spectrum. It is common to record the solar emission as a function of both frequency and time because the solar radio flux at different frequencies has a higher degree of correlation with all other solar phenomena. Thus, solar radio emissions are a very useful indicator of major solar activity, such as sunspot cycle or magnetic reversal cycle \citep{Chandra2011, Li2012, Xie2017}. The radio flux emissions from Sun changes gradually in intensity. The emission of solar radio flux probably originates from three different sources, namely, either from the undisturbed solar surface or from initially active regions, and also from short-lived enhancements above its daily level. The dependency of spectral lines on various solar atmospheric regions has been studied mostly by analyzing the time series of each individual spectral line, emanating from various regions. Each region of the solar atmosphere could be a source of radio flux at a definite frequency, and time-series of such emission could be useful for spectral analysis \citep{Kane2004, Kane2009}. 

The radio emission at various discrete frequencies originates from different heights in the solar chromosphere, transition region and inner and outer corona. Various models proposed on plasma density and radio emissions provide approximate heights of its origin as a function of emissions frequency. \citet{Aschwanden1995} and \citet{Melendez1999} presented a generally accepted model that predicted plasma density at various heights above the solar surface. Using the model they proposed, the plasma frequency at various heights can be calculated. We know that the radio emissions having frequencies less than or equal to the plasma frequency will be absorbed, but radiation with a second harmonic (at twice the plasma frequency) will propagate \citep{Vats2001}. According to the plasma density models suggested, it is found that usually higher frequencies can be supposed to be emanating from deeper regions, i.e., from lower solar altitudes. Besides the plasma density model, the temperature-height profiles given by \citet{Fontenla1999, Vats1998, Vats2001} can also be used to determine the height from where the radio flux at particular frequencies originates in the corona. 

Solar radio fluxes emanated at different heights are also used to find out the temporal and spatial change of coronal rotation \citep{Vats1998, Vats2001, Chandra2009, Chandra2011, Li2012, Xie2017, Bhatt2017, Bhatt2018}. The rotation of the solar corona can also be used as an indicator of the differential rotation of subphotosperic layers. \citet{Vats2001}, uses disk-integrated simultaneous measurements of solar radio flux for the frequency range of 275-2800 MHz and showed a strong dependence of the coronal rotation period on the heights in the corona, i.e., the sidereal coronal rotation period increases with an increase in height in the solar atmosphere. While \citet{Bhatt2017} contradict the result by using the same data set and obtained that solar corona rotates slower at higher altitudes. By analyzing the coronal rotation period as a function of latitude between $\pm60\circ$ for the period 1999-2001, \citet{Chandra2009} obtained that as compared to the photosphere and chromosphere, solar corona rotates less differentially. Similar results were confirmed by \citet{Chandra2010} who evaluated the differential rotation of soft X-ray corona obtained from the solar full-disk images of a soft X-ray telescope. \citet{Chandra2011} uses 2.8 GHz emitted solar radio flux data for the time span 1947-2009 and reported that sidereal coronal rotation period varies from 19.0 days to 29.5 days with an average of 24.3 days, and a 22-year component (which might be related to the 22-year Hale magnetic cycle or magnetic field reversal) exists in the long-term rotation. According to \citet{Xu2020}, variations in the lengths of the solar atmospheric rotational period indicate several harmonics of the solar cycle period, which is modulated by the solar activity cycle.
 
\begin{figure}
		\includegraphics[width=80mm]{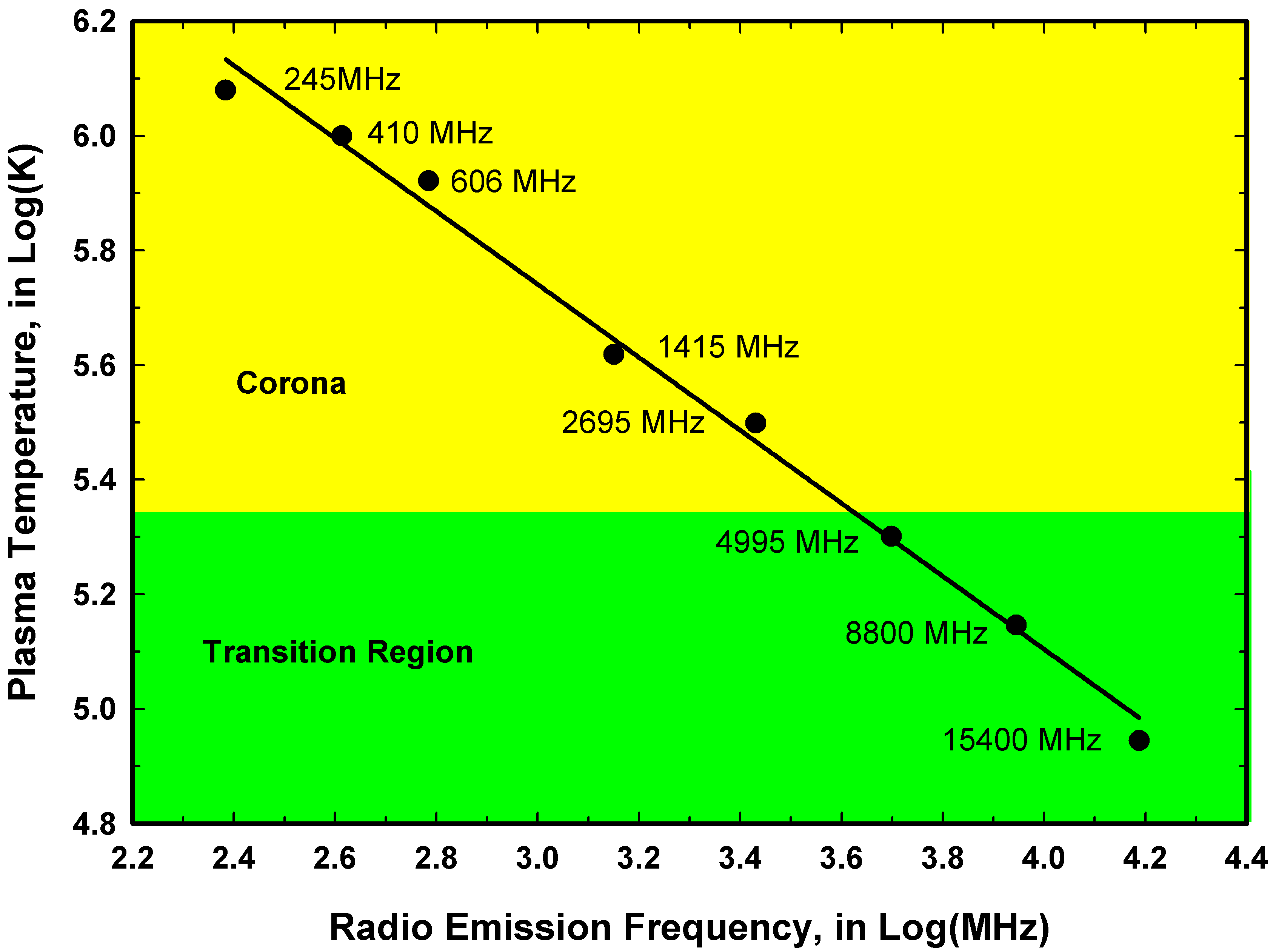}
    \caption{Temperature-frequency profile of solar atmosphere (Green colour - Transition region and Yellow colour - Corona); emission frequencies and temperatures are in deg. K and in MHz, respectively (after \citet{Kane2009}).}
    \label{fig:Figure 1}
\end{figure}

\begin{figure*}
  \centering{
  \includegraphics[width=155mm]{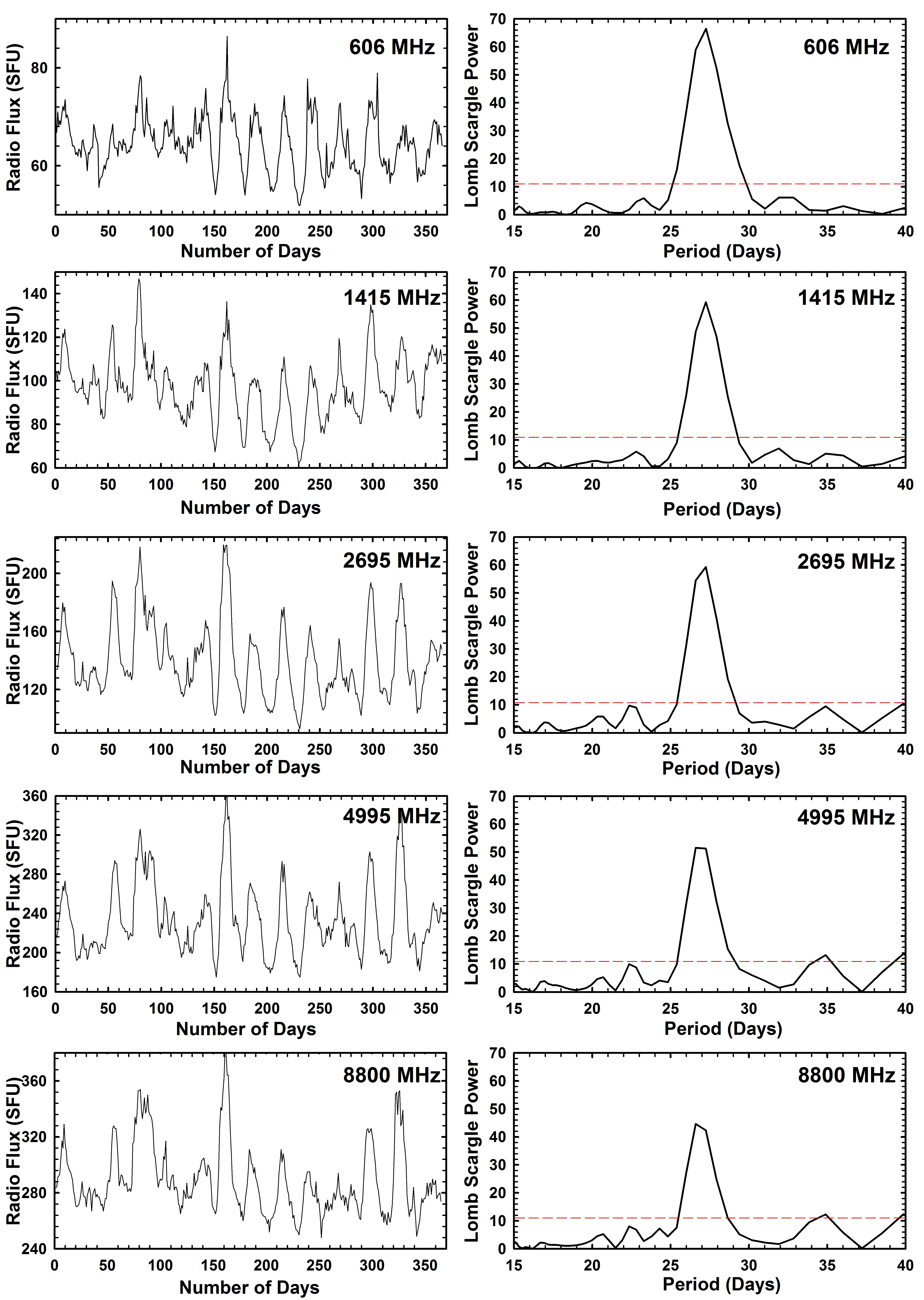}}
 \caption{Left Panels: The radio emission at 606, 1415, 2695, 4995 and 8800 MHz frequency is plotted against days for the year 1969. Right Panels: The LSP of respective radio emission is plotted against period (FAP = 0.05).}
 \label{fig:Figure 2}
 \end{figure*}

In the present work flux modulation method is used to determine the rotation period present in the time series of solar radio flux. The data sets used for the study were available for a span of 44 years (1967-2010) at eight frequencies 245, 410, 606, 1415, 2695, 4995, 8800 and 15,400 MHz, which originating from different height from the solar atmosphere. The radio flux in the 245-15400 MHz range is most likely escaping from the upper corona to the upper chromosphere. \citep{Kane2009}. The height at which the radio flux at a particular frequency emitted is estimated from the temperature-height profiles (shown in Figure~1) and found that flux of 245 MHz emitted at the height of $\sim$ 12,000 km, 410 MHz at $\sim$ 10,000 km, 606 MHz at $\sim$4,000 km, 1415 MHz at $\sim$2,500 km, 2695 MHz at $\sim$2,800 km approx), 4995 MHz from outer transition region at $\sim$2,200 km, 8800 MHz from mid-transition region and 15400 MHz from inner transition region at $\sim$2,100 km above photosphere in the solar atmosphere \citep{Kane2004, Kane2005, Kane2009}.

The fractal dimension are sometimes used as an index to quantify the time variability of radio emission. For the years 1976–1990, \citet{Watari1996} measured the fractal dimensions of solar radio fluxes data at 245, 410, 610, 1415, 2695, 2800, 4995, 8800, and 15400 MHz. The sunspot number and radio emission around 3 GHz have identical annual variations in fractal dimensions, implying a close relationship between them.

According to the height-frequency profile of the plasma and gyromagnetic frequencies given by \citet{Gary2004, Raulin2005}, for free-free opacity equal to unity ($f_{\tau}=1$), the highest frequency at which plasma radio emission may have been detected in the solar corona is about 8000 MHz. Thus, emissions between frequencies 245-4995 MHz originates in the lower corona (less than $0.2R_{\odot}$) whereas emissions at frequencies 8800 and 15400 MHz originate in the transition region. According to the \citet{Gary2004} (based on Figure 4.1 given in the paper), we learn that the plasma and bremsstr$\ddot{a}$hlung emission, both can coexist for the fluxes at 245, 410 and 606 MHz. Whereas, only bremsstr$\ddot{a}$hlung free-free emission will prevail for the fluxes at 1415 and 2695 MHz. The gyro-emission mechanism will dominate over free-free emission for the frequencies at 8800 and 15400 MHz.

The flux from active regions is normally dominated by gyro-resonance emission, while plage regions displaced from sunspots emit thermal bremsstr$\ddot{a}$hlung emission. As the number of sunspots changes during the solar cycle, the contribution of the rotationally modulated portion should change as well. \citet{Schmahl1994} discussed that the fraction of total flux at intermediate frequencies (usually in the 1000-9400 MHz range) is much greater than the fraction at lower and higher frequencies in the most of the solar active years. As a result, the fraction spectrum peaks similarly to the gyro-resonance spectrum, indicating that in active years, the free-free contribution is a smaller fraction of the modulated portion. The lower frequency fraction, on the other hand, increases to match the intermediate frequency fraction at solar minimum. As a result, at solar minimum, the rotationally-modulated component's spectrum is flatter, suggesting a greater contribution from plage-associated sources.

The slowly-varying components, or S-components, are those fraction of radio emissions in the centimeter (15400, 8800, 4995, and 2695 MHz) and decimeter (1415, 606, and 410 MHz) wavelength range whose variation shows the rotational modulation of long-lived active regions. It's either thermal bremsstr$\ddot{a}$hlung from hot and dense coronal condensations in active regions, which are correlated with plages, or gyroresonance emission whether it's originate above sunspots or intense magnetic fields. Around 1000-2000 MHz, plage-associated emission persists, while sunspot-associated emission prevails above 5000 MHz. Between 2000 and 5000 MHz, plage‐associated emission and spot‐associated emission, both may coexists \citep{Schmahl1994, Schmahl1995, Watari1996}.

High altitudes in the corona are the region from where the quiet-Sun emission originates at meter (245 MHz) and decimeter (1415, 606, and 410 MHz) wavelengths. Whereas, contributions from the chromosphere dominate the quiet-Sun emission at centimeter (15400, 8800, 4995, and 2695 MHz) wavelengths. But at decimeter wavelengths, both the chromosphere and the corona lead to the quiet-Sun emission \citep{Watari1996}.

Estimated heights of origin of the flux at frequencies 245, 410, 606, 1415, 2695, and 4995 MHz through the model suggested by \citet{Gary2004} are therefore, $\sim$1,90,000, $\sim$1,10,000, $\sim$53,000, $\sim$12,000, $\sim$5,100 and $\sim$2,900 km, respectively (all in lower solar corona) and $\sim$2,200 km (in upper transition region) for 8800 MHz and $\sim$2,100 km (in lower transition region) for 15,400 MHz radio flux.

Different heights in the solar corona can emit radio emission at specific frequencies. Radio emissions with frequencies less than or equal to the plasma frequency are found to be absorbed, whereas second harmonic (at twice the plasma frequency) radiation propagates. As a result, an electron density model can be used to measure the location of the solar radio emission \citep{Vats2001}. \citet{Zucca2014} has obtained the electron density as a function of height, from $R_{\odot}$ to $2.5R_{\odot}$, for the active region, quiet Sun, and coronal hole using radial profiles of electron density. The plasma frequency ($\nu_p$) and the electron density ($n_e$) are related through the following expression \citet{Gary2004},

\begin{equation}
			\nu_p = 8.98 \times 10^3 \sqrt{n_e}
\end{equation}

whereas, the gyro-frequency ($\nu_B$), is expressed as
	
\begin{equation}		
			\nu_B = 2.6 \times 10^6 B
\end{equation}

where $B$ is the magnetic field strength in Gauss.

Even with numerous proposed height-frequency models, the exact height ranges from which radio frequencies actually escape are still unknown, and all height estimates are roughly correct. However, it is generally agreed in most proposed models that emissions at higher frequencies will escape from lower depths in a relative sense (lower solar altitudes). As a result, the qualitative comparison should be appropriate, particularly when the data used are averages over long periods of time.

\section{Data Analysis and Methodology}

The disk-integrated solar radio emissions observed at different frequencies (245-15400 MHz) for each day are recorded from May 1966 to December 1987 at Sagamore Solar Hill observatory, Massachusetts, USA  and from Jan 1988 to December 2010 at other observatories (Palehua, Hawaii; San Vito, Italy; Learmonth, Australia; and also at Sagamore Hill, USA) \citep{Kane2009}. The noon flux monitoring at 606, 1415, 2695, 4995, and 8800 MHz were first launched by the Sagamore Hill Solar Radio Observatory in 1966. In 1967, the observational frequency was augmented to 15400 MHz, then to 242 MHz in early 1969, and finally to 410 MHz in early 1971. The noon flux disk-integrated data recorded daily are available in the public domain through National Centers for Environmental Information (NCEI-NOAA). To investigate the altitudinal variation in solar rotation, these time series of disk-integrated radio flux emanates at a particular frequency are taken from the year 1967-2010, each of which belongs to a particular height in the solar transition region or lower corona as shown in Figure~1. An annual time series are then generated for each of the frequencies (245, 410, 606, 1415, 2695, 4995, 8800 and 15,400 MHz) for the span of the year 1967 to 2010.

\subsection{Data Analysis}

After the initial review, it is found that the radio frequency data of 15,400 MHz originates from transition region have data gap more than permissible for statistical analysis. Hence, we discarded the 15,400 MHz data and therefore, we were not able to estimate any results from this region. Furthermore, perhaps due to unsymmetrical structure in the higher corona region, from where the radio frequency at 245 and 410 MHz originates, the variation in the emission lacks any systematic rotational modulation which ultimately prevent us to estimate the rotation period of this region. \citet{Bhatt2018} also reported the similar observation for fluxes at 245 and 15,400 MHz. The turbulence effect is inversely proportional to the increasing frequency \citep{Vats2001}. It may be possible that due to the turbulence in the intervening media from the Sun to the radio telescope the emissions at 245 and 410 MHz gets more randomized but other flux at higher frequencies propagates unattenuated due to turbulence.

By determining the fractal dimension of a time series of radio flux, the irregular aspects of solar radio flux variations can be quantified. The specified time series has a fractal dimension of 1, indicating that it is completely regular. As the time series' variance becomes more irregular, the fractal dimension increases. After calculating fractal dimensions for each year and each frequency \citet{Watari1996} found that, for frequency ranges less than 1000 MHz and frequencies greater than 10,000 MHz, the fractal dimension has a high value of 1.7. Around 3 GHz, the fractal dimension reaches a minimum. The fractal dimension is greater than 1.5 at 245, 8800, and 15,400 MHz. However, for frequencies between 610 and 4995 MHz, the fractal dimension is less than 1.3 and reaches a minimum around 3 GHz. This means that there is a regular systematic rotational modulation in the solar atmosphere for the time series of radio flux at frequencies between 610 and 4995 MHz \citep{Watari1996}.

\citet{Bhatt2018} also investigated fractal dimension of 410, 1415, 2695, 4995, and 8800 MHz solar radio flux measurements of two solar activity cycles recorded on a daily basis and shown that with increasing frequency, fractal dimension increases, implying that randomness increases towards the inner corona. Their research also reveals that solar activity has a greater impact on low frequencies, while solar activity has a lesser impact on higher frequencies because they found that at lower frequency fractal dimension difference between solar maximum and solar minimum is much more compared to the difference at higher frequency.

\citet{Watari1996} reported that in the distribution of radio-source lifetimes simulated at 1.5 GHz, larger fractal dimensions were correlated with shorter e-folding time constants. According to \citet{Gary2004}, The cyclotron frequency and its harmonics are represented by the gyro-emission. The gyro-emission at the cyclotron frequency ($\nu = \nu_B$) is normally not relevant for the solar corona within active regions, rather, the emission typically occurs at the third harmonic, i.e., at ($\nu = 3\nu_B$). The $3\nu_B$ line stretches from 1–2 GHz just above $f_{\tau}=1$ level to 20 GHz below the $f_{\tau}=1$ level. This is why, the gyro-emission dominates predominantly over bremsstr$\ddot{a}$hlung free-free emission for radio emission $\sim$20 GHz, like 15.4 GHz emission.

According to the characteristic frequencies to height in the solar atmosphere model, suggested by \citet{Gary2004}, the plasma emission dominates at frequencies ranging from 30 kHz to several hundred MHz. Plasma emission dominates over thermal free-free emission in this frequency spectrum primarily because the plasma level is above the $f_{\tau}=1$ level at these frequencies. While plasma levels in active regions are usually below $f_{\tau}=1$ at decimeter frequencies range (200 MHz–1 GHz), even though the plasma emission may still be significant over free-free emission due to coronal inhomogeneity in this height range. This may lead to the unsystematic variability found in the data of radio flux at 245 and 410 MHz.

The rotational modulation will take place on every emission (irrespective of the emission process) from the Sun. However, for estimating the rotation period from these observations require the following conditions to be fulfilled; (a) emission should originate from certain region/s of the Sun and not uniformly from the entire Sun, and (b) emission should not be too much modified by the irregular scattering in the intervening medium (from source to receiver).

The outcome of \citet{Watari1996, Vats2001, Gary2004, Bhatt2018} justifies the use of annual time series of radio frequency 606, 1415, 2695, 4995 and 8800 MHz. For the reasons stated therein and above, the metric emission at 245 and 410 MHz and centimeter emission at 15,400 MHz radio data found to have more noise and hence excluded for further study. The rotation period of inner corona and transition region, from where the radio frequency 606, 1415, 2695, 4995 and 8800 MHz originates, could have been estimated successfully with few data gaps years. A typical examples of such annual time series of disk-integrated flux of year 1967 are plotted against the days of year in Figure~2 (left panel) for each of the frequencies under investigation. The intensity variation shows the dominating periodically varying component throughout the period of study (annual data set in our case). But, it is also found that in some cases the varying component is quite weak and random during some part under study. This may be because of the presence of short-lived solar features like radio noise, radio burst, etc.. Such time regions in the time series have to be discarded from the time series under study.

\begin{figure}
		\includegraphics[width=85mm]{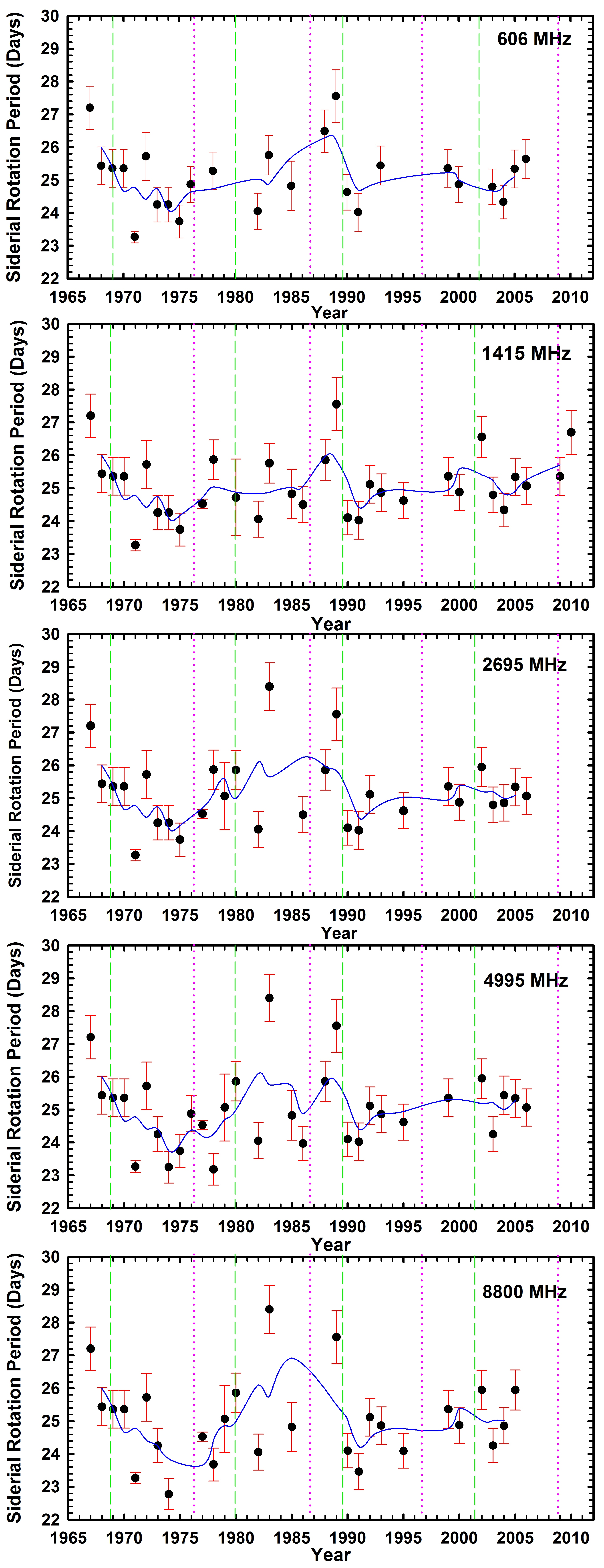}
    \caption{Scattered plot shows sidereal rotation period (SRP) estimated between year 1967-2010 at five radio frequencies and smooth curves are three years smoothed SRP. The vertical dotted (pink) and dashed (green) lines represents the positions of the years of solar minima and maxima, respectively.}
    \label{fig:Figure 3}
\end{figure}

\begin{figure}
		\includegraphics[width=87mm]{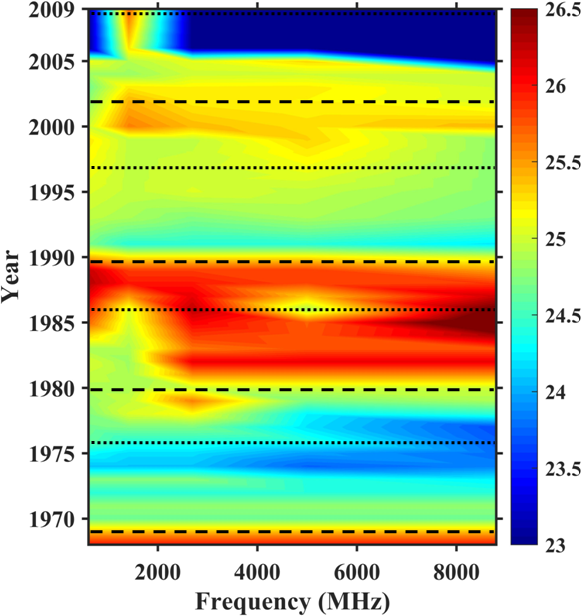}
    \caption{Space-time diagram of sidereal rotation period (3-yrs smoothed) obtained from solar emissions at five radio frequency 606-8800 MHz observed during 1967-2010. The colour codes (shown on top) represent the sidereal rotation period, which ranges from 24.0 to 26.5 days. The horizontal dotted and dashed lines represents the positions of solar minima and maxima years, respectively.}
    \label{fig:Figure 4}
\end{figure} 

\subsection{Methodology}

\citet{Vats1998} found that the time series of radio frequency at 2.8 GHz shows particularly significant modulation owing to coronal rotation, indicating that radio emissions are less impacted by irregularities and unsystematic features inside the solar corona. Hence, time series of such radio flux/radiation can be used to estimate rotation period of solar corona. In the present work, flux modulation is used to estimate the rotation period by radiation/flux emitted from the Sun at different frequencies. This method has been employed with the disk-integrated simultaneous daily measurements of solar flux at different frequencies (such as radio flux 2.8 GHz) to measure differential rotation as a function of altitude in the solar atmosphere \citep{Vats1998, Vats2001, Chandra2011, Li2012, Xie2017, Bhatt2017}. 

The radio flux can itself illustrate a rotational modulation (as shown in left panels of Figure~2) in its variation. But, determining rotation period directly from modulated variation in radio flux will not be possible directly and hence some time series statistical analysis technique is required to find statistically plausible period of rotation. The rotation periods were determined mostly using auto-correlation analysis or continuous wavelet transformation analysis in previous works \citep{Vats2001, Chandra2011, Li2012, Xie2017}. But in the present work we employed the Lomb-Scargle Periodogram (LSP) to get probable rotational periodicity present in the multi-frequency time series of radio emissions.

This type of periodogram is used for a frequency/period analysis of such data which are not collected at a regular time interval or have any data missing in time-series \citep{Lomb1976, Scargle1982, Horne1986, VanderPlas2018}. The Lomb-Scargle periodogram is widely used in the field of astrophysics to lessen aliasing from unequally spaced data. Unlike Fast Fourier Transformation (FFT), LSP also gives accurate calculation of a spectrum estimator from unevenly sampled time-series and is analogous to least-square fitting of sinusoidal curves, i.e., it searches the sinusoidal periodicity and find out the best fit of a sinusoidal function, as compared to the time-series, for a selected period or frequency. The location of dominant peaks of the exponential probability distribution function of power \textsl{P(T)} on the period of oscillation, \textit{T} gives rotation periods of all periodic components present in that time series. After observing power spectrum of periodogram, FAP (false alarm probability) measurement is essential to verify the probability that the dominating peak is due to any true periodic signal and is not the outcome of any noise present randomly in the time series. The FAP Assessment of a FAP level of power numerically requires at least the computation of 10/power \citep{Chowdhury2013, Delisle2020, Mancuso2020}.

\section{Results and Discussion}

Our investigation depends on picking out any statistically significant periodic signals present in the multi-frequency time series of radio flux observed during year 1968-2009 at Sagamore Solar Hill observatory, Massachusetts, USA. The Lomb-Scargle periodogram (LSP), along with the auto-correlation function has widely been used for detecting periodicity in radio flux time series \citep{Vats1998, Vats2001, Chandra2011, Li2012, Xie2017, Bhatt2017}.

\subsection{Spatial variation of solar rotation with height}

The right panels of Figure~2 shows the periodogram power plotted as a function of the period of rotation for five observed radio frequencies at 606, 1415, 2695, 4995 and 8800 MHz for the year 1969. The plots clearly shows the statistically significant prominent peaks power are well above the $95\%$ confidence level (FAP) and which shows that the peaks are due to presence of true periodic signal and are not because of any inherent noise in the radio data. The synodic rotation period can be determined by single out the position of statistically significant prominent peak on the axis of period of rotation. The initial investigation shows that the locus at which the peaks are situated in the plots of Figure~2 (left panel) are almost same for all the frequencies under study, which means that the period of rotation are invariant with respect to the frequency. Such invariance have been found in most of the years during 1967-2010, with few exceptions. The emissions at these frequencies emanate from different heights in inner solar corona and transition region. So, our investigation shows that the rotation of inner solar corona, as well as transition region, is almost rigid with respect to altitude above solar surface \citep{Chandra2009, Morgan2011}. 

The result differs completely with the study reported by \citet{Vats2001} and \citet{Bhatt2017}, who used approximately two years radio flux data observed at Cracow Astronomical Observatory, Poland. According to \citet{Vats2001} the coronal rotation period increases with the frequency of emissions (275-1755 MHz) emanate from the solar corona, whereas \citet{Bhatt2017} reported with the same data that period of rotation decreases with increase in frequency of the coronal radio flux. The differential rotation of solar inner corona with respect to the height in solar atmosphere reported by \citet{Vats2001} and \citet{Bhatt2017} is completely ruled out in our study.

The observed almost rigid rotation with respect to altitude in transition region and inner corona may be understand considering the similar quasi-rigid rotation behaviour reported with low latitude coronal holes. Coronal holes are cool, low-density regions on the Sun that can be found at low latitudes as well as near the polar caps \citep{Raju2005}. Many coronal holes have a surprisingly rigid rotation, indicating that field-line reconnection happens regularly in the corona. Magnetic interactions between active regions and coronal holes have been investigated as a potential explanation for coronal holes' rigid rotation. Magnetic reconnection between open and closed magnetic fields (known as interchange reconnection) must occur at the boundary of coronal holes, contributing to the rigid rotation of coronal holes. Magnetic reconnection at the coronal holes boundaries (CHBs) appears to be the cause of rigid rotation, as reported in many observations \citep{Kahler2002, Madjarska2004, Raju2005, Hiremath2013}. 

So we can conclude that, our and others observations suggest that the large-scale features of corona rotates more rigidly than the underlying photosphere in general, and that the rotation becomes more rigid as the height is increased up to about $2.5R_{\odot}$, where the plasma is dominated by the magnetic field. The rapid decrease of higher order multipoles of the magnetic field above the photosphere can be attributed to this nature \citep{Antonucci2020}.

\begin{figure}
		\includegraphics[width=85mm]{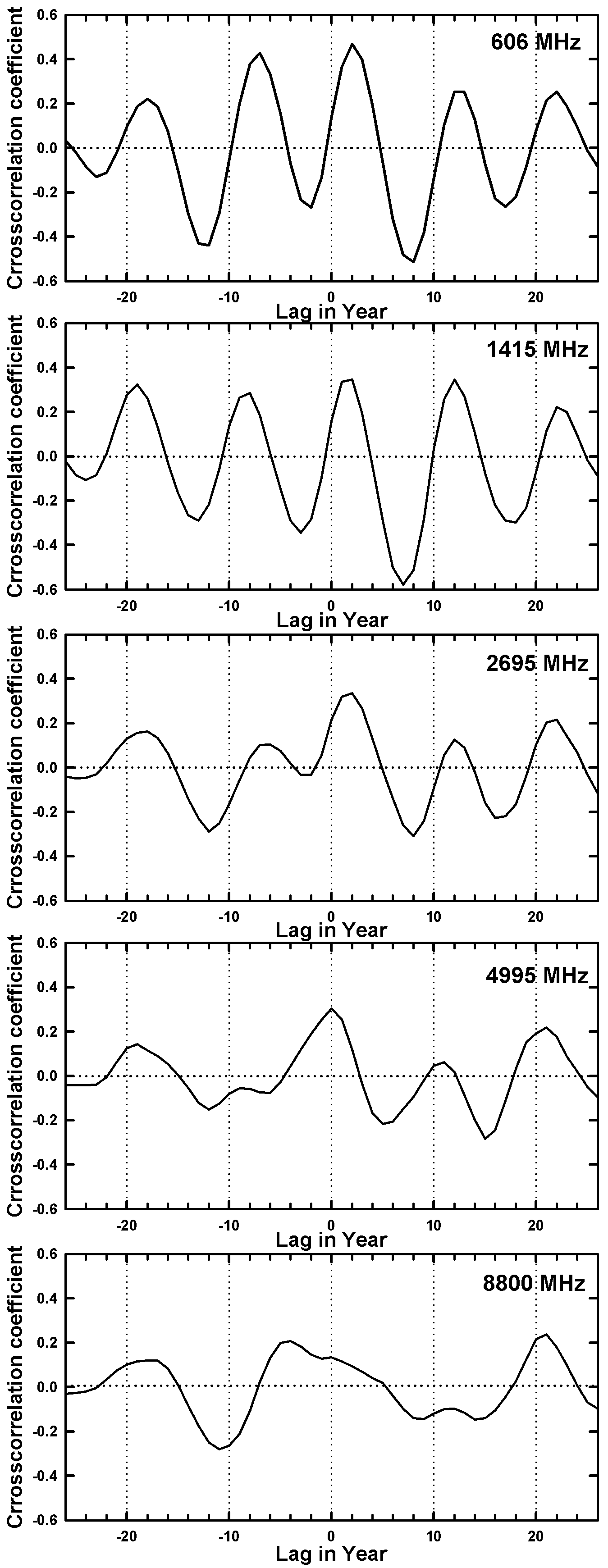}
    \caption{Cross-correlograms are plotted between two time-series, one is sunspot numbers (SSN) and other is 3-yr smoothed sidereal rotation period (SRP), at radio frequency 606, 1415, \& 2695 MHz (emission from inner corona) and 4995 \& 8800 MHz (emission from transition region).}
    \label{fig:Figure 5}
\end{figure}

\subsection{Temporal variation of solar rotation with years}

The synodic rotation period, $T$ obtain through LSP analysis (as shown in left panels of Figure~2) is actually apparant rotation period. The Sun rotates little extra due to the simultaneous rotation of Earth along with Sun during each of its rotation. So, we have to find the actual period known as sidereal rotation period, $T_{o}$, which can simply be calculated using the well known formula \citep{Chandra2009, Deng2020}, 

\begin{equation}
			T_{o}=\frac{\left(T\times 365.26\right)}{\left(T+365.26\right) }
\end{equation}

where, $T$ is the synodic rotation period.

The sidereal rotation period calculated, by the formula given above, for the radio emission at five discreet frequencies obtain from each annual time series between the years 1967-2010 are plotted in five panels and shown in Figure~3 as scatter plots along with their statistical errors. The years of sunspot cycle maxima (by dashed vertical lines) and years of sunspot cycle minima (by dotted vertical lines) are also shown in each panels of Figure~3 \citep{Chandra2011, Xie2017}.

The scatter plots in five panels of Figure~3, shows the temporal variation in sidereal rotation period of five particular layers of solar lower corona and transition region. The variation in rotation period seems quite random in nature, at first sight, unless they are smoothed by taking running mean of 3 years.  The 3-yrs. smoothed rotation period data for each frequencies are also plotted along with the scatter plots of rotation in Figure~3. These smoothed curves of rotation period show the presence of longer duration periodic components into it, which seems to wax and wane along with the sunspot cycle. The curves shows maxima around two alternate solar maximum years, i.e., around year 1969 and 1990. The findings are quite similar to the study investigated with 2800 MHz radio data by \citet{Chandra2011, Xie2017}. It is therefore pertinent to analyze further the interrelationship between the solar activity cycles and rotation period of inner corona and transition region.

\subsection{Space-Time variation in sidereal rotation period}

The smoothed sidereal rotation period (by taking 3-yrs running mean) shown as smooth curves in plots of Figure~3 are again plotted in Figure 4 as space-time diagram for the multi-frequencies (606-8800 MHz) flux emitted during the years 1968-2009. The smoothed period varies between 24.0 - 26.5 days, which have been depicted through the six different colour codes, from violet to red, and separated by 0.5 days from each other. The contour plot is overlaid by horizontal lines, dotted and dashed, which represents the years of solar activity minima and maxima falls during the years of observation, i.e., 1968-2009. 

The diagram along with dotted lines of solar minima clearly show that the rotation period changes its wax and wane at the dawn of any solar cycle. For example, in 1976 and 1997 (beginning of cycle-21 and -23, respectively) the rotation periods touches its least value across the frequencies, which originates from different layers of solar atmosphere, whereas in 1986, when cycle-22 kicks off, the rotation periods of lower corona and transition region are almost at its peak. It is evident from the Figure 4, that rotation periods rises from one solar minima to other consecutive minima, and then falls till next solar minima. This systematic rise and fall in sidereal rotation periods have been followed by all the five frequencies, although the contrast in highest and lowest value of period is more for those frequencies, which emerges from transition region, in comparison to that originates from lower corona. The wax and wane of sidereal rotation period prominently shown in Figure 4 suggests the linkage of rotation period with the phases of 11-yrs and 22-yrs solar cycles.

\begin{figure}
		\includegraphics[width=82mm]{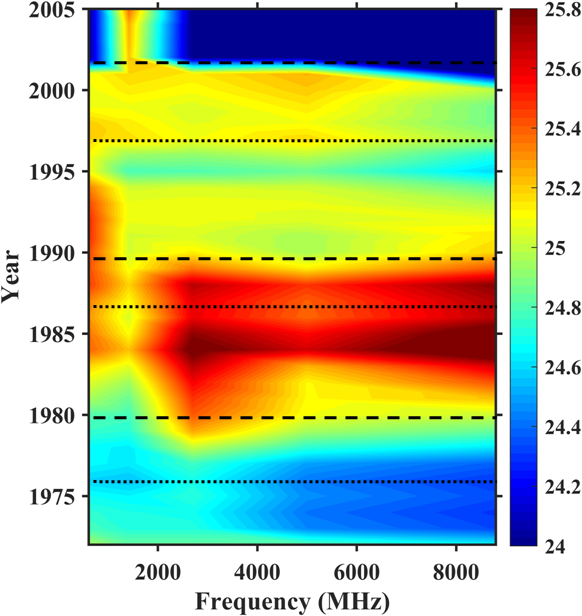}
    \caption{Space-time diagram of sidereal rotation period (11-yrs smoothed) obtained from solar emissions at five radio frequency 606-8800 MHz observed during 1972-2001. The colour codes (shown on top) represent the sidereal rotation period, which ranges from 24.4 to 25.8 days. The horizontal dotted and dashed lines represents the positions of solar minima and maxima years, respectively.}
    \label{fig:Figure 6}
\end{figure}

\subsection{Correlation with solar activity cycles}

To investigate it further we performed cross-correlation (CC) analysis between sidereal rotation period and annual sunspot numbers after taking the 3-years running mean of both the time series. Although, cross-correlation are used to determine the degree of fit between two time series of same data size. But, the CC analysis can also be used to see the mutual correlation between solar rotation and other observed solar indices \citep{Chandra2011, Xu2020}.

The cross-correlograms between smoothed sidereal rotation period and sunspot numbers for five different frequencies 606, 1415, 2695, 4995 and 8800 MHz. are plotted in Figure 5. The curves plotted are smooth, cyclic and prominent in most of the cases. They rises and falls periodically with almost same time period for the emissions originate from inner corona, but shows ambiguity in some parts in the curves belongs to transition region. Even-though, the consecutive maximas or minimas of cross-correlation coefficient are almost 11 - 13 years apart from each other for all the five frequencies. The smoothness of curve, in some cases, is affected by the presence of gap of data of some year's. So we can conclude, that when shorter duration components (less than 3 year) are suppressed than the periodic components of $\sim$11 yrs, which are also embedded in the analyzed rotation period data and their presence is perhaps due to the 11 yrs sunspot activity cycle, which can be seen in the temporal variation of rotation period of any solar layer \citep{Chandra2011, Li2012, Mei2018, Xu2020}.

Taking the lead from Figure 3 and to find the presence of periodic components of longer duration, the evaluated rotation period profiles for all five radio emissions are once again smoothed by taking 11 years running mean. Such smoothing has eliminated all those periodic components which have periodicity shorter than $\sim$11 yrs. After smoothing, the rotation periods are plotted again in space-time diagram (as shown in Figure 6) for the observation years from 1972 to 2001. Due to smoothing up to 11-yrs, the rotation period varies less in comparison to raw data or 3-yrs smoothed data (as shown in Figure 3 and 4). It ranges between 24.4 - 25.8 days, which are colour coded in Figure 6 as violet to red colour. The space-time diagram in Figure 6 confirms the presence of 22-yrs cycle in sidereal rotation period of lower corona and transition region. The horizontal dotted lines drawn over the contour diagram reveals that any solar cycle begins either when the rotation periods of solar atmospheric layer are at its minimum (as evident in solar cycle 21 \& 23) or at maximum (as in solar cycle 22) of its value. The colour distribution for solar cycle 21 and 22 appears like a mirror image of each other. This variation is quite similar to the behaviour of solar magnetic field reversal phenomenon which happens after every $\sim$22 years. In the same way, our investigation establishes that the rotation period of the layers under study repeats its successive minimum or maximum period after every $\sim$22 years \citep{Chandra2011, Xie2017, Xu2020}.

\begin{figure}
		\includegraphics[width=85mm]{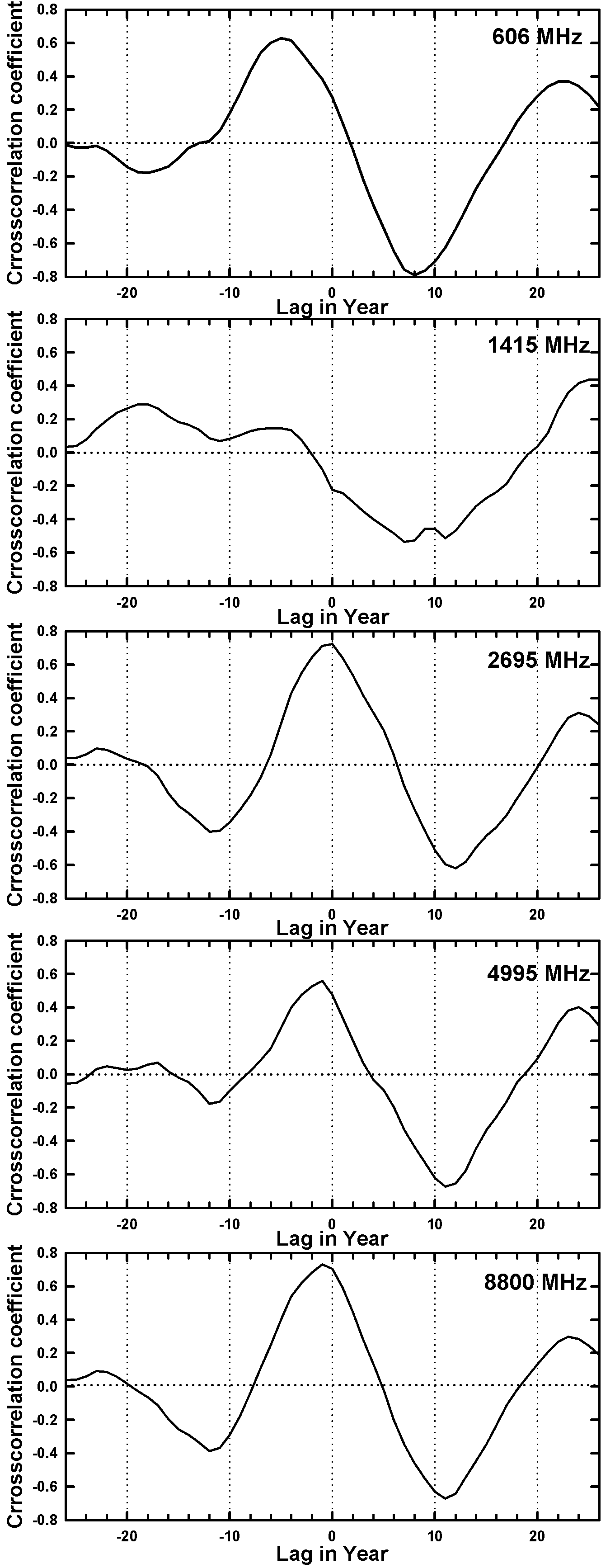}
    \caption{Cross-correlograms are plotted between two time-series, one is sunspot numbers (SSN) and other is 11-yr smoothed sidereal rotation period (SRP), at radio frequency 606, 1415, \& 2695 MHz (emission from inner corona) and 4995 \& 8800 MHz (emission from transition region).}
    \label{fig:Figure 7}
\end{figure}

The 11-yrs smoothed rotation period time series are cross-correlated with the similarly smoothed annual sunspot numbers (SSN). The cross-correlograms, so obtained, are plotted for each frequency in Figure 7. The CC coefficients are prominent at peaks and valley and curves shows almost cyclic behaviour in its variation. Although the curves in some cases are not so smooth (like in case of 1415 MHz emission), even though the difference between two consecutive peaks and valley of cross-correlation coefficient have years difference of almost 22-26 yrs. This peak to peak or valley to valley difference may be because of the presence of longer duration periodic components in rotation period which may have the some influence on 22 years magnetic field reversal cycle \citep{Chandra2011, Deng2020, Xu2020}.

As we have seen that the variations of the rotation period of the layers of inner corona and transition region obtain through the different radio wavelengths are in tandem with the solar cycle. There are similar views about the rotation profiles of coronal holes. \citet{Navarro-Peralta1994} and references therein, reported that the rotation profile of coronal holes is also dependent on the phase of the solar cycle. As we know that the Sun is characterized by coronal holes near two polar caps with opposite magnetic polarities at the minimum of the solar activity cycle. There are two types of coronal holes found in the mid-latitudes. Coronal holes that are disconnected from polar coronal holes are often found near active regions and have an incidence rate that closely corresponds to the solar activity cycle. The equatorial extension of a polar coronal holes is the second type of coronal holes. During times other than solar maximum, however, their rate of incidence is reasonably consistent. These rotate nearly rigidly, as contrasted to disconnected coronal holes, which rotate differentially with latitude, although slightly more rigidly than the photospheric differential rotation rate \citep {Navarro-Peralta1994, Insley1995, Wang1996, Kahler2002, Prabhu2018}.

\subsection{Interchangeability of rotation speed of transition region and corona with solar cycles}

The contours of 3 yrs and 11 yrs running mean (Figure 4 and 6, respectively) show that there is a significant 22 year periodicity in the temporal variation of sidereal rotation periods. In order to see this feature even more clearly,  we plot the average 11 yrs running mean of (a) all the five frequencies; (b) two frequencies of transition region (4995 and 8800 MHz), and (c) three coronal emission frequencies (606, 1415 and 2695 MHz). These three averages are shown in Figure 8. All three curves show a clear 22 yrs periodicity. Additionally, this variation also show an interesting outcome as follows:

(1)	During 1972 (or even before) to 1977  corona rotates faster than transition region.

(2)	During 1978 to 1988 transition region rotates faster than corona.

(3)	During 1989 to 2000 corona rotates faster than transition region.

(4) Both rotate at the same speed in the years 1979, 1989 \& 2000 (almost at solar maximas).

The curves show a tendency of similar change over before 1972 and also after 2000. This is a new characteristic of rotational profile in the solar atmosphere; that corona rotates faster or slower than transition region in different epoch. This also has $\sim$22 years periodicity.

\section{Conclusion}

We investigated multi-frequency yearly time series of radio flux observed at five frequency 606, 1415, 2695, 4995 and 8800 MHz during year 1967-2010 at Sagamore Solar Hill Observatory, USA (Figure 1). The continuous temporal variation in synodic rotation period obtains through Lomb-Scargle Periodogram (LSP) analysis for each frequency during study period confirms the presence of prominent random component in the solar rotation (Figure 2). When scattered sidereal rotation period are smoothed to suppress the random component by taking 3-yr running mean, it hints the presence of longer duration periodic cycle, perhaps $\sim$11-yr Schwabe cycle \citep{Mei2018} and $\sim$22 yrs Hale cycle \citep{Xu2020} (Figure 3).

The space-time diagram in Figure 4 and 6, are the smoothed sidereal rotation period (by taking 3-yrs and 11-yrs running mean) plotted against radio emissions at five different frequencies for the period under study. The diagram reveals that any solar cycle kicks off at the inflexion point of rotation period of solar layer. In these diagrams, we can see that the solar cycle 21 \& 23 starts when a particular solar layer rotates fastest and the cycle 22 begins when it rotates at its slowest pace. Such behaviour seems to be in phase with the 22-year magnetic field reversal cycle \citep{Chandra2011, Li2012, Xie2017}.

To see any relation with the phases of 11 yrs solar cycle, the cross-correlation (CC) analysis between 3-yr running mean of sidereal rotation period and sunspot number (SSN) has been performed. The CC analysis confirms the presence of $\sim$11yr periodic component in the obtained rotation period. The position of peaks in cross-correlograms (Figure 5) show that the rotation period of inner corona and sunspot number are in-phase but in transition region, it is leading slightly ($\sim$1 year).

The presence of all the component having periodicity less than 11 yrs has been eliminated by taking 11-yrs running mean. When time series of 11-yrs running mean of sidereal rotation period are cross-correlated with time series of annual sunspot numbers, the cross-correlograms show peaks at interval of almost 22 years. The peaks for rotation of inner corona is lagging for $\sim$3 yrs, whereas for upper transition layer, it is almost in phase with the sunspot cycle \citep{Chandra2011, Deng2020}.

When, average of 11 yrs running mean rotation period, which emanates from corona is compared with those emanates from transition region (as shown in Figure 8), the variation in rotation period of corona and transition region show the prominent presence of 22-yrs Hale's cycle. Besides that, the curves show that the rotation period of corona is more than transition region near the dawn of solar cycle 21 \& 23, whereas it is less almost at the start of cycle 22. It means corona rotates faster in comparison to transition region in alternate solar cycle, and $\it{vice-versa}$. It is evident from Figure 4, that their speeds interchange its track at the time of solar maxima. Thus, we can also conclude that swapping of rotation speed of corona and transition region essentially follows the 22-yrs periodicity. 

Thus, the period of rotation for five radio frequencies at 606, 1415, 2695, 4995 and 8800 MHz originates from lower corona and upper transition layer during the year 1967-2010 shows the presence of 11-yrs and 22-yrs component besides the dominating random components.

\begin{figure}
		\includegraphics[width=85mm]{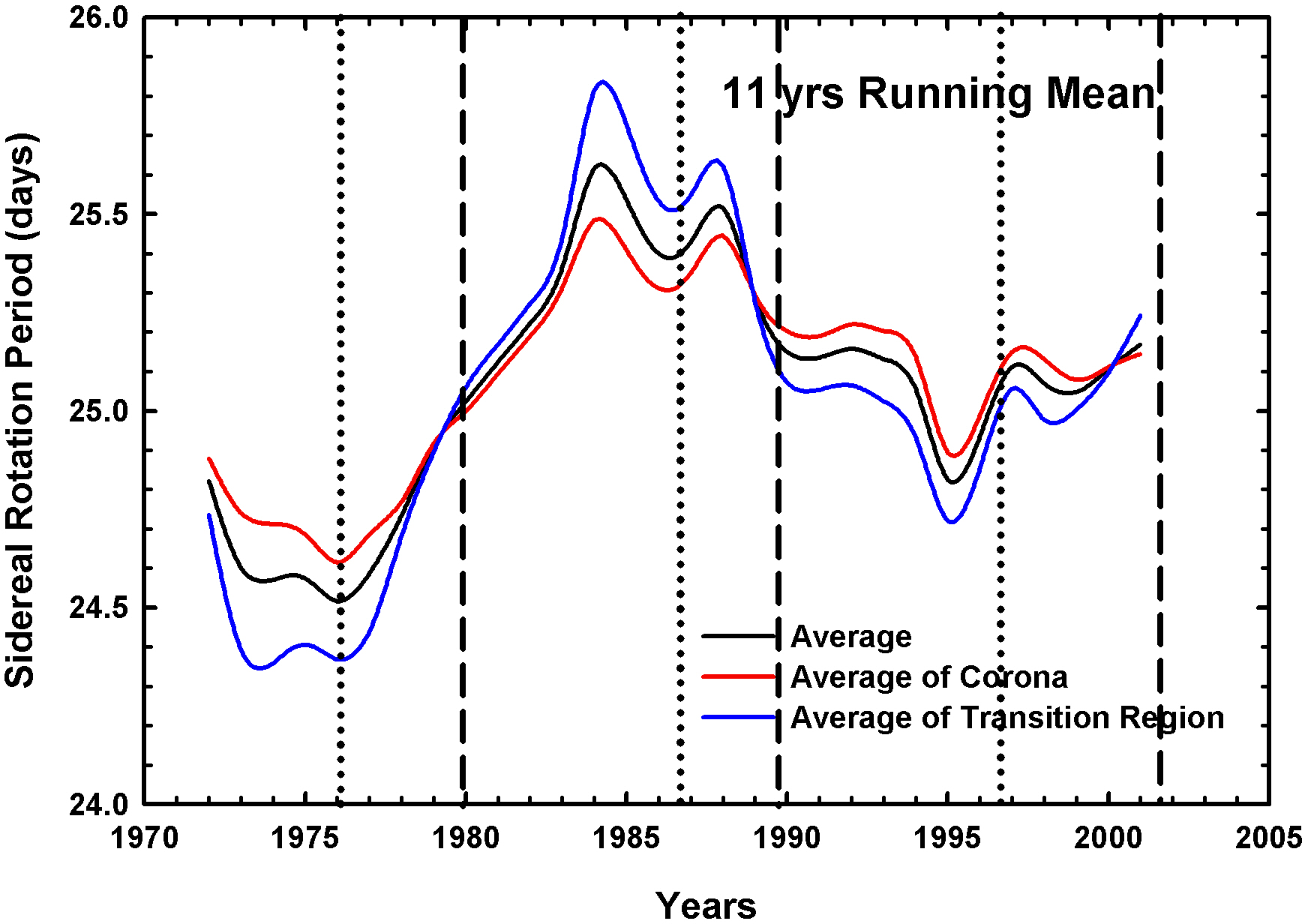}
    \caption{11 yrs running mean of average sidereal rotation period (SRP) estimated for all the five radio emissions (as black line) plotted against years is compared with average SRP of three emissions originated from corona (red line) and two emisions from transition region (blue line). The vertical dotted and dashed lines represents the positions of solar minima and maxima years, respectively.}
    \label{fig:Figure 8}
\end{figure}

\section*{Acknowledgments}

The authors wish to acknowledge the numerous observers and data archivers of the radio flux and sunspot data used in this present work. The sunspot number data are credited to the source; WDC-SILSO data, Royal Observatory of Belgium, Brussels and the radio data to the source; National Centers for Environmental Information (NCEI-NOAA), USA. This research was supported by the CSJM University, Kanpur, India under its financial research grant scheme. We appreciate the anonymous reviewers' thorough reading of our manuscript as well as their numerous insightful comments and suggestions.

\section*{Data Availability}

The data-sets used in this present work were derived from the sources freely available as public data. The data underlying this article are available at {\it National Centers for Environmental Information (NCEI-NOAA)} at \url{https://www.ngdc.noaa.gov/stp/space-weather/solar-data/solar-features/solar-radio/noontime-flux/sagamore-hill/} and {\it World Data Center - SILSO, Royal Observatory of Belgium, Brussels} at \url{http://www.sidc.be/silso/datafiles}.


\bsp

\label{lastpage}

\end{document}